\documentclass[12pt]{article}
\usepackage{array}
\usepackage{graphicx}
\usepackage{amssymb}
\usepackage{amsmath}
\usepackage{multirow}
\input{epsf}
\usepackage{cite}
\def\@fmsl@sh#1#2#3{\m@th\ooalign{$\hfil#1\mkern#2/\hfil$\crcr$#1#3$}}
 \def\eq#1\en{\begin{equation}#1\end{equation}}
\def\s[#1,#2]{[#1\stackrel{\star}{,}#2]}
\def\sx[#1,#2]{[#1\stackrel{\star_{x}}{,}#2]}

\newcommand{\nc}{\newcommand}
\nc{\beq}{\begin{equation}}
\nc{\eeq}{\end{equation}}
\nc{\beqa}{\begin{eqnarray}}
\nc{\eeqa}{\end{eqnarray}}

\def\bc{\begin{center}}
\def\ec{\end{center}}

\def\to{\rightarrow}

\def\gsim{\mathrel{\mathpalette\atversim>}}

\def\bc{\begin{center}}
\def\ec{\end{center}}

\def\gsim{\mathrel{\rlap{\lower4pt\hbox{\hskip1pt$\sim$}}

    \raise1pt\hbox{$>$}}}       %greater than or approx. symbol

\def\gsim{\mathrel{\rlap{\lower4pt\hbox{\hskip1pt$\sim$}}
    \raise1pt\hbox{$>$}}}       %greater than or approx. symbol

\begin{document}
\makeatletter
\def\fmslash{\@ifnextchar[{\fmsl@sh}{\fmsl@sh[0mu]}}
\def\fmsl@sh[#1]#2{%
  \mathchoice
    {\@fmsl@sh\displaystyle{#1}{#2}}%
    {\@fmsl@sh\textstyle{#1}{#2}}%
    {\@fmsl@sh\scriptstyle{#1}{#2}}%
    {\@fmsl@sh\scriptscriptstyle{#1}{#2}}}
\def\@fmsl@sh#1#2#3{\m@th\ooalign{$\hfil#1\mkern#2/\hfil$\crcr$#1#3$}}
\makeatother
%\baselineskip 24pt

%%%%%%%%%%%%%%%%%%%%%%%%%%%%%%%%%%%%%%%%%%%%%%%%%%%%%%%%%%%%%%%%%
%%%
%%%                      TITLE PAGE
%%%
%%%%%%%%%%%%%%%%%%%%%%%%%%%%%%%%%%%%%%%%%%%%%%%%%%%%%%%%%%%%%%%%%
\thispagestyle{empty}
\begin{titlepage}
\boldmath
\begin{center}
  \Large {\bf An alternative view on the electroweak interactions}
    \end{center}
\unboldmath
\vspace{0.2cm}
\begin{center}
{ {\large Xavier Calmet}\footnote{x.calmet@sussex.ac.uk}
}
 \end{center}
\begin{center}
{\sl Physics and Astronomy, 
University of Sussex,   Falmer, Brighton, BN1 9QH, UK 
}
\end{center}
\vspace{\fill}
\begin{abstract}
\noindent
We discuss an alternative to the Higgs mechanism which leads to gauge invariant masses for the electroweak bosons. The key idea is to reformulate the gauge invariance principle which, instead of being applied as usual at the level of the action, is applied at the level of the quantum fields. In other words, we define gauge invariant quantum fields which are used to build the action.  In that framework,  the Higgs field is not necessarily a physical degree of freedom but can merely be  a dressing field that does not propagate. If the Higgs boson is not propagating, the weak interactions must become strongly coupled below 1 TeV and  have a non-trivial fixed point and would thus be renormalizable at the non-perturbative level. On the other hand, if a gauge invariant Higgs boson is introduced in the model, its couplings to the fermions and the electroweak bosons can be quite different from those expected in the standard model.
\end{abstract}  
\end{titlepage}

%\pacs{}

%%%%%%%%%%%%%%%%%%%%%%%%%%%%%%%%%%%%%%%%%%%%%%%%%%%%%%%%%%%%%%%%
%%%
%%%                     INTRODUCTION
%%%
%%%%%%%%%%%%%%%%%%%%%%%%%%%%%%%%%%%%%%%%%%%%%%%%%%%%%%%%%%%%%%%%

\newpage

Modern theories of particle physics are based on the concept of gauge invariance \cite{Yang:1954ek}. The action which describes fully the model is built using fields which transform according to chosen representation of the gauge group.  Besides being a Lorentz scalar, the action is built in such a way that it is invariant  under gauge transformations. This approach has been extremely successful and has led to a realistic model of particle physics, the standard model, which has survived for over 40 years now. The only remaining sectors to probe of the standard model are the Higgs sector and Yukawa sector which respectively give masses to the electroweak gauge bosons and the fermions of the model. However, the Higgs mechanism \cite{Higgs:1964pj} is the weakest feature of the model since it requires the introduction of a fundamental scalar field, the elusive Higgs boson, which is the last particle to be discovered and the source of the hierarchy problem of the standard model. 

In this paper we propose an alternative viewpoint on  gauge invariance which opens up the possibility to formulate mass terms for gauge bosons in a manifestly gauge invariant way without introducing a propagating fundamental scalar degree of freedom. The intuition comes from the St\" uckelberg action for a massive $U(1)$ photon \cite{Stueckelberg:1900zz}:
\begin{eqnarray}
S=\int d^4x \left  (-\frac{1}{4} F_{\mu\nu} F^{\mu\nu}+\frac{1}{2} (\partial_\mu \phi + m A_\mu) (\partial^\mu \phi + m A^\mu) \right ),
\end{eqnarray}
where $A_\mu$ is the photon, $\phi$ is a real scalar field which takes on values in a real 1 dimensional affine representation of $\mathbb{R}$ which is the Lie algebra of $U(1)$. This action is invariant under local $U(1)$ transformations of the type $\delta A_\mu = \partial_\mu \epsilon $ and $\delta \phi = -m \epsilon$. In other words, this action is gauge invariant despite having a massive gauge boson. The Proca action is recovered by choosing the gauge $\phi=0$. This mechanism is an alternative to the Higgs mechanism for a $U(1)$ gauge invariant theory. It has been applied to extensions of the  standard model \cite{Kors:2004dx}. Unfortunately it does not generalize to $SU(N)$ gauge symmetries, since the non-Abelian St\" uckelberg action leads to a violation of unitarity already at the tree-level, see \cite{Kors:2005uz} for a recent review. As we shall see, our alternative to the Higgs mechanism has the same problem with perturbative unitarity and renormalization in the classical sense of the term. 

In order to illustrate the new mechanism, let us first reconsider an old idea that goes back to Dirac \cite{Dirac:1955uv} (see also \cite{Bagan:1999jf,Ilderton:2010tf} for more recent work in this direction), namely, that electrons and photons, since they are observable objects, should be gauge invariant. Classically, an action such as the one describing quantum electrodynamics is built by requiring that it would be invariant under gauge transformations and Lorentz invariant. One obtains: 
\begin{eqnarray}
S=\int d^4x \left (\bar \psi(x) (\gamma^\mu i D_\mu- m) \psi(x) -\frac{1}{4} F^{\mu\nu}(x)F_{\mu\nu}(x) \right),
\end{eqnarray}
where $D_\mu=\partial_\mu - i e A_\mu(x)$ is the covariant derivative, $e$ is the electric charge, $F_{\mu\nu}(x)=\partial_\mu A_\nu(x) - \partial_\nu A_\mu(x)$, $\gamma_\mu$ are the Dirac matrices,  $\psi(x)$ is the quantum field which describe the electron and the positron, $\bar \psi(x) = \psi^\dagger(x) \gamma_0$ and $A_\mu(x)$ is the electromagnetic potential which is related to the electric and magnetic fields.  
By construction, this action is invariant under local $U(1)$ gauge transformations
 \begin{eqnarray}
 \psi^\prime(x)&=& \exp(-i \alpha(x)) \psi(x) \\
 A_\mu^\prime(x)&=&  A_\mu(x) -\frac{1}{e}  \partial_\mu \alpha(x). 
 \end{eqnarray}
 On the other hand, the field which represents the electron  and the positron is not gauge invariant. However, only gauge invariant quantities are observable. This was a motivation for Dirac to formulate quantum electrodynamics using dressed fields \cite{Dirac:1955uv}. He introduced the dressed fermion field:
 \begin{eqnarray}
\underline \psi (x) = \Omega_D^\star \psi(x)
\end{eqnarray}
with 
\begin{eqnarray} 
\Omega_D = \exp \left ( - i e \int d^4z f^\mu(x-z) A_\mu(z) \right ), 
 \end{eqnarray}
where $A_\mu$ is the electromagnetic four-vector which represents the photon field.
Note that $\underline \psi(x) $ is locally gauge invariant  as long as $\partial_\mu f^\mu(z)=\delta^{(4)}(z)$, but it is not invariant under global $U(1)$ transformations. 
Inspired by the St\" uckelberg mechanism, instead of using $\Omega_D$ we will use 
\begin{eqnarray} 
\Omega = \exp \left ( i e \sigma(x)  \right ), 
 \end{eqnarray}
with $\delta \sigma(x) = - \alpha(x)/e$ under a $U(1)$ gauge transformation. This dressing  has the benefit of being local. We define
\begin{eqnarray}
\underline \psi (x) = \Omega^\star \psi(x).
\end{eqnarray}
We can also define a gauge invariant electromagnetic field
\begin{eqnarray} 
{\underline A}_\mu(x)  &=&\frac{i}{2 e} \Omega^\star \stackrel{\leftrightarrow}D_\mu \Omega= A_\mu(x) - \partial_\mu \sigma(x),
\end{eqnarray}
where $D_\mu=\partial_\mu - i e A_\mu(x)$ and $ \Omega^\star\stackrel{\leftrightarrow}D_\mu \Omega= \Omega^\star D_\mu \Omega  -  \Omega (D_\mu \Omega)^\star$.

 One can then reformulate the action of quantum electrodynamics in terms of the gauge invariant fields
\begin{eqnarray}
S=\int d^4x \left (\bar{\underline \psi}(x) (\gamma^\mu i \underline D_\mu- m)  \underline \psi(x) -\frac{1}{4}  \underline F^{\mu\nu}(x)  \underline F_{\mu\nu}(x) \right ),
\end{eqnarray}
where $\underline D_\mu=\partial_\mu - i e \underline A_\mu(x)$  and , $\underline F_{\mu\nu}(x)=\partial_\mu \underline A_\nu(x) - \partial_\nu \underline A_\mu(x)$.
Note that the underlying gauge principle prevents the coupling of  neutral particles such as the neutrino to the photon, indeed it is not possible to define a gauge invariant field if the underlying fermion is not transforming under the underlying local gauge symmetry.
The remarkable feature of this alternative gauge principle is that one can write down a gauge invariant mass term for the photon:
\begin{eqnarray}
S_{\mbox{photon mass}} =\int d^4 x \frac{1}{2}m_A^2 \underline A_\mu(x)  \underline A^\mu(x).
\end{eqnarray}
Note that this is essentially a reformulation of the St\" uckelberg mechanism in terms of gauge invariant fields, however, the gauge principle is applied in a different way. Instead of introducing the gauge fixing fields in the action, they are introduced in the fundamental degrees of freedom of the action. 

This formalism can easily be extended to the standard model with the usual gauge group $SU(3) \times SU(2)_L \times U(1)$, but let us first consider $SU(2)_L \times U(1)$ for the sake of simplicity. Our goal is to consider an alternative to the Higgs mechanism. The idea consists in promoting the usual $SU(2)_L$ valued fields of the Standard Model to $SU(2)_L$ invariant fields. The gauge symmetries of the model are $SU(2)_L  \times U(1)_Y$. Furthermore, there is a global $SU(2)$ symmetry. We consider a theory with exactly the same particle content as that of the standard model. The hypercharge assignments are the same as well. We shall see that we do not need to introduce a propagating Higgs boson.

Let us start by introducing a scalar $SU(2)_L$ doublet 
\begin{eqnarray}
\phi=\left(\begin{array}{c}  \phi_1 \\
   \phi_2
  \end{array}
\right ).
\end{eqnarray}
In the sequel, this field, which in the usual standard model would be the Higgs field, will play the role of a dressing field. One can now define 
\begin{eqnarray}
\Omega_2=\frac{1}{\sqrt{\phi^\dagger
    \phi}}\left(\begin{array}{cc}  \phi_2^* & \phi_1 
    \\ -\phi_1^* & \phi_2
  \end{array}
\right )
\end{eqnarray}
and 
\begin{eqnarray} 
\Omega_Y =  \exp \left (   i Y e \sigma(x)  \right ),
 \end{eqnarray}
where $Y$ is the hypercharge of the particle under consideration. 
These matrices can be used to define gauge invariant fields
\begin{eqnarray}
\underline \psi^a_L&=&  \Omega_Y^\star \Omega_2^\dagger \psi^a_L
\end{eqnarray}
for the left-handed fermions and 
\begin{eqnarray}
\underline W^i_\mu&=& \frac{i}{2g} \mbox{Tr} \ \Omega_2^\dagger \stackrel{\leftrightarrow}
{D_\mu} \Omega_2 \tau^i
\end{eqnarray}
for the electroweak bosons which are not gauge bosons in the usual sense since they are gauge invariant. The right-handed fermions are given by
\begin{eqnarray}
\underline \psi^a_R&=&  \Omega_Y^\star \psi^a_R.
\end{eqnarray}
The gauge invariant hyperphoton is defined by
\begin{eqnarray} 
{\underline {\cal A}}_\mu(x)  &=&\frac{i}{2  e} \Omega^\star \stackrel{\leftrightarrow}{\cal D}_\mu \Omega
\end{eqnarray}
with ${\cal D}_\mu=\partial_\mu - i  e {\cal A}_\mu(x)$.
 
One can easily construct the following gauge invariant kinetic terms
\begin{eqnarray}
-\frac{1}{2} \ \mbox{Tr} \ {\underline F}_{\mu \nu} {\underline F}^{\mu \nu} -\frac{1}{4}  {\underline f}_{\mu \nu} {\underline f}^{\mu \nu} 
\end{eqnarray}
with
\begin{eqnarray} \label{g}
({\underline D}_\mu)&=& \Omega_2^\dagger (D_\mu) \Omega_2=\partial_\mu - i g \underline W_\mu
\\ \nonumber
{\underline F}_{\mu \nu}&=&\frac{i}{g}[{\underline D}_\mu,{\underline D}_\nu],
\end{eqnarray}
and ${\underline f}_{\mu \nu} = \partial_\mu \underline {\cal A}_\nu -\partial_\nu \underline {\cal A}_\mu$. Note that the $SU(2)$ part of the dressing is that considered by 't Hooft and many others\cite{tHooft:1998pk,'tHooft:1980xb,Banks:1979fi,Dimopoulos:1980hn,Mack:1977xu,Visnjic:1987pj,class1,Nyffeler:1999ap,Calmet:2000th,Calmet:2001rp,Calmet:2002mf}. It is important to realize that the coupling constant $g$ in equation (\ref{g}) is the coupling constant which appears in $D_\mu$. This is a consequence of the local  $SU(2)$ gauge invariance. It implies that the different coupling constants between the weak bosons and the different fermions are fixed by the gauge coupling  and are identical to those of the standard model.

So far, there was no difference between the usual electroweak interactions and the alternative formulation of the model based on gauge invariant quantum fields. However, there is a new feature which does not exist in the usual standard model. We can use the gauge invariant formulation of the quantum fields to introduce a gauge invariant mass terms for the hyperphoton  and the weak bosons $W^i$: 
\begin{eqnarray} 
\frac{m^2}{8} \ \mbox{Tr} \left ( (a \underline W_\mu^i \tau_i + b\underline {\cal A}_\mu \tau_0)
(a \underline W^{\mu j} \tau_j + b\underline {\cal A}^\mu \tau_0) \right ), 
\end{eqnarray}
where $\tau_0$ is the $SU(2)$ identity matrix. After the diagonalization of  the mass matrix we  find the physical fields
\begin{eqnarray} 
\underline W^\pm_\mu &=& \frac{\underline W^1_\mu \mp i  \underline W^2_\mu }{\sqrt{2}}, \\ 
\underline Z_\mu &=& \frac{ a \underline W^3_\mu - b  \underline {\cal A}_\mu }{\sqrt{a^2+b^2}}, \\
\underline A_\mu &=& \frac{ a \underline W^3_\mu + b  \underline {\cal A}_\mu }{\sqrt{a^2+b^2}}.
\end{eqnarray}
Let us now introduce $\sin \theta_W=b/\sqrt{a^2+b^2}$ which as in the standard model is a free parameter. We have the following mass spectrum
\begin{eqnarray} 
m_{W}=  \frac{a}{2} m,  \   m_Z=m_{W} \cos  \theta_W  \ \mbox{and} \ m_A=0. 
\end{eqnarray}
The nonlinear couplings of the electroweak bosons are thus the same as those of the standard model. Let us emphasize that the electroweak sector is that of the standard model without a propagating Higgs. As in the case of the St\" uckelberg mechanism, this alternative to the Higgs mechanism suffers from a unitarity problem at tree-level in the scattering of electroweak bosons. We shall come back to this after the discussion of fermion masses.
Let us first point out that we could trivially reproduce the full standard model by introducing a gauge invariant Higgs boson
\begin{eqnarray} 
\underline \phi= \Omega_Y^\star \Omega_2^\dagger \phi.
\end{eqnarray}
We could thus introduce a propagating degree of freedom in the scattering of electroweak bosons and restore tree-level unitarity in the usual way with a gauge invariant Higgs boson. A  potential for the gauge invariant Higgs boson and Yukawa couplings would generate as in the usual standard model the masses of the electroweak bosons and of the fermions. The connection between this approach to the electroweak interactions to the standard model in the unitarity gauge is made by picking the gauge $U=\Omega_2^\dagger(\phi)$ with $\phi= {\bf 1} \underline \phi$.

Let us now return to the non-propagating Higgs model. Fermion masses can be introduced in a gauge invariant way as well:
\begin{eqnarray} 
\bar{\underline \psi}_L^a M \underline \psi_R^a + h.c. 
\end{eqnarray}
where $M= diag(m_{up}, m_{down})$. Note that right-handed fermions are singlets under the local $SU(2)$ gauge symmetry but doublets under the global one. Let us now discuss the inclusion of $SU(3)$. So far we focussed on leptons but quarks can be included in the same way. Gauge invariant dressed quark fields have been introduced in \cite{Lavelle:1995ty}. Thus, including quarks  does not represent any further difficulty. We can simply multiply our $SU(2)_L\times U(1)$ gauge invariant fields by a further dressing field $\Omega_3$ introduced in  \cite{Lavelle:1995ty} which takes care of the  local $SU(3)$ invariance. The CKM matrix is generated as in the standard model.

Let us now come back to the question of unitarity. The results presented in this paper lead to some interesting insight in the Higgs mechanism. The Higgs mechanism is not needed as such to break the gauge symmetry since gauge invariant mass terms can be explicitly constructed, however it is needed to restore perturbative unitarity in the scattering of electroweak bosons. Gauge theories with massive gauge bosons are only consistent if the masses are generated via spontaneous symmetry breaking. This conclusion has been reached a long time ago using different arguments \cite{Cornwall:1974km,LLS,Lee:1977yc,Lee:1977eg}. However, tree-level unitarity could be restored by different means.  If the  scalar field is not propagating, the electroweak interactions have to become strongly coupled around 1 TeV thereby restoring unitarity. If this non-perturbative theory has a non-trivial fixed point, it would be renormalizable at the non-perturbative level, as conjectured by Weinberg for quantum general relativity \cite{fixedpoint}. This possibility is very attractive, since it solves the fine-tuning problem of the standard model by eliminating the Higgs boson as a fundamental particle.

As mentioned above, we could still introduce a scalar degree of freedom, i.e. a gauge invariant Higgs boson.  The Higgs boson does not need to have a vacuum expectation value. It might or might not contribute to the masses of the electroweak bosons and the fermions.  It is interesting to note that in this case, there are potentially two contributions to the electroweak gauge bosons and fermion masses. If we introduce a gauge invariant Higgs boson and generate at least part of the masses of fermions and electroweak bosons through the gauge invariant mechanism proposed above, the production and decay modes of the Higgs boson could be quite distinct from that of the standard model, as for example studied in \cite{Calmet:2000vx}.  The Yukawa coupling are essentially arbitrary if the masses of the fermions are simply gauge invariant terms. However, tree-level unitarity would have to be restored in this sector as well \cite{Maltoni:2000iq,Maltoni:2001dc,Dicus:2004rg,Appelquist:1987cf}. Note that if the gauge invariant Higgs boson resolves the tree-level unitarity problem in the scattering of electroweak bosons, it also needs to generate the totality of their masses as in the standard model.

We shall now consider the couplings of the gauge invariant Higgs boson to the remaining particles in more detail. Let us assume that the Higgs boson has a vacuum expectation $v^\prime$ which does not need to be equal to $246$ GeV since part of the masses of the electroweak bosons could be generated by our new mechanism. In the sequel $v^\prime$ is a free parameter. It is easy to derive the Feynman rules for the gauge invariant Higgs. We have the following vertices
\begin{eqnarray} 
W^+_\mu W^-_\nu H &\to& \frac{-i g^2 v^\prime}{2} g_{\mu\nu}, \nonumber \\
Z_\mu Z_\nu H &\to& \frac{-i g^2 v^\prime}{2 \cos^2\theta_W} g_{\mu\nu}, \nonumber \\
H H H &\to& - 6 i |\lambda| v^\prime, \nonumber \\
H H H H &\to& -6 i |\lambda|,  \nonumber \\
W^+_\mu W^-_\nu H H  &\to& \frac{-i g^2}{2} g_{\mu \nu},  \nonumber \\
Z_\mu Z_\nu H  H &\to&  \frac{-i g^2}{2 \cos^2 \theta_W} g_{\mu \nu},  \nonumber \\
\bar f_i f_i H  &\to& \frac{-i}{\sqrt{2}} G_i. \nonumber 
\end{eqnarray}
The gauge coupling $g$ and the weak mixing angle $\theta_W$ are the same as in the usual standard model. On the other hand $v^\prime$ and the Yukawa couplings $G_i$ are free parameter which are not necessarily related to the masses of the electroweak bosons or corresponding fermions. We denote the self-Higgs coupling as usual by $\lambda$. Note that the strength of the  two Higgs bosons two electroweak bosons interaction is fixed by the underlying gauge invariance. If a gauge invariant Higgs boson is introduced, its pair production must be as in the standard model. However, the coupling of a single Higgs to two electroweak bosons or two fermions is not fixed by gauge invariance or the mass of the particles and is essentially arbitrary. The decay widths for the decay of the gauge invariant Higgs boson to fermions are of the same form as in the standard model $\Gamma(H \to \bar f_i f_i)=G^2_i M_H/(16 \pi)$, but can be very different since they depend on the unconstrained Yukawa couplings $G_i$. The same holds for the decays to electroweak bosons. If $v^\prime$ is not equal to the usual vacuum expectation of the standard model Higgs boson, a mechanism is needed to solve the unitarity problem.

Another application for gauge invariant fields is that it opens up the possibility of coupling the standard model particles to a hidden sector (for example describing dark matter) in a  richer way than just the Higgs portal or the kinetic term of the hyperphoton, see, e.g., \cite{Calmet:2009uz} and references therein. We now have fermion or electroweak boson portals of the type  $\underline{ \bar \psi}_L\  \underline \psi_R {\cal }O_{hidden,1}$ or $\underline W^i_\mu \underline W^{i \mu} O_{hidden,2}$ where $O_{hidden,i}$ are operators from the hidden sector. Our mechanism opens up the door to new approaches in  model building. It is easy to extend these considerations to supersymmetric models \cite{Calmet:2000ws}.

To summarize, we have considered an alternative application of the Yang-Mills local gauge principle. This enable us to formulate gauge invariant masses for the electroweak bosons and the fermions. The real difficulty in the electroweak interactions is not the generation of  masses for the particles of the model, but rather to do so in a way which preserves tree-level unitarity. In our formulation of the electroweak interactions, tree-level unitarity could be restored by strongly interacting electroweak bosons in the TeV region if we do not introduce a gauge invariant Higgs boson. If  we introduce a propagating  gauge invariant scalar field, the couplings of the Higgs boson to the fermions are not necessarily uniquely determined by the masses of the given particles and the production cross-sections and decay widths can be quite different from the standard model expectations. Unless the gauge invariant Higgs boson vacuum expectation value generates all the masses of the particles of the model, a new mechanism needs to be implemented to address the tree-level unitarity problem. Let us stress finally that we believe that the most interesting feature of this alternative perspective to the electroweak interactions is the possibility to remove the Higgs boson of the particle spectrum and thus to solve the hierarchy problem. There is a clear experimental signature for this mechanism, namely strongly interacting electroweak bosons in the TeV region.

\bigskip

{\it Acknowledgments:}
I would like to thank  Sebastian J\" ager and Oliver Rosten for helpful discussions.
This work is supported in part by the European Cooperation in Science and Technology (COST) action MP0905 ``Black Holes in a Violent Universe". 

%\newpage

%%%%%%%%%%%%%%%%%%%%%%%%%%%%%%%%%%%%%%%%%%%%%%%%%%%%%%%%%%%%%%%%%
%%%
%%%                     BIBLIOGRAPHY
%%%
%%%%%%%%%%%%%%%%%%%%%%%%%%%%%%%%%%%%%%%%%%%%%%%%%%%%%%%%%%%%%%%%%

\bigskip

%\newpage
%\vskip .75 in

\end{document}